\documentclass[a4paper,11pt]{article}
\usepackage{jcappub} 
\usepackage{lineno}
\usepackage{graphicx}
\usepackage{subcaption}
\usepackage{makecell}
\usepackage{multirow}
\usepackage{siunitx}
\arxivnumber{2601.22340}
\title{\boldmath Grey-body factors of higher dimensional regular black holes in quasi-topological theories}

\author{Juan Pablo Arbelaez}
\affiliation{Centro de Matemática, Computação e Cognição (CMCC), Universidade Federal do ABC (UFABC),
Rua Abolição, CEP: 09210-180, Santo André, SP, Brazil}

\emailAdd{juan.arbelaez@ufabc.edu.br}

\abstract{We study grey-body factors and Hawking radiation of higher-dimensional regular black holes arising in quasi-topological gravity. These spacetimes incorporate infinite–curvature corrections that remove the central singularity while preserving an event horizon and a well-defined semiclassical description. We show that, for all considered regular black hole models, the transmission of radiation and the corresponding Hawking evaporation are systematically  suppressed compared to the singular black hole solutions of General Relativity.}

\begin{document}
\maketitle
\flushbottom

\section{Introduction}

Black holes are not perfect black bodies. Quantum mechanically, they emit radiation with a thermal spectrum modulated by frequency-dependent transmission coefficients, known as grey-body factors, which encode the interaction of the emitted quanta with the curved spacetime geometry outside the event horizon. These factors determine both the energy emission rates and the detailed spectral properties of Hawking radiation \cite{Hawking:1975vcx}, thereby providing a direct link between semiclassical quantum effects and the classical scattering problem for perturbations in black hole backgrounds. As such, grey-body factors play a central role in understanding black hole evaporation, stability, and potential observational signatures of modifications to General Relativity \cite{Page:1976df,Page:1976ki,Kanti:2004nr}.

In recent years, regular black holes -- spacetimes free of curvature singularities -- have attracted considerable attention as viable effective descriptions of quantum-corrected gravitational collapse. In many constructions, regularity is achieved either through effective matter sources or via higher-curvature corrections, while preserving the existence of an event horizon. From a phenomenological perspective, such models provide a setting in which quantum-gravity-inspired modifications can be explored without encountering pathological divergences at the classical level. However, regularity alone does not guarantee that the dynamical, thermodynamical, or radiative properties of these black holes closely resemble those of their singular counterparts. This makes a detailed analysis of wave propagation, scattering, and Hawking emission indispensable. Consequently, a substantial body of literature has been devoted to the study of perturbations, quasinormal spectra, grey-body factors, and Hawking radiation of various regular black hole models \cite{Simpson:2021biv,Lopez:2022uie,DuttaRoy:2022ytr,Meng:2022oxg,Li:2022kch,Konoplya:2023aph,Konoplya:2023ppx,Myrzakulov:2023rkr,Filho:2023voz,Bolokhov:2023ozp,Al-Badawi:2023lvx,Konoplya:2023ahd,Konoplya:2023bpf,Jha:2023wzo,Balart:2023odm,Huang:2023aet,Bolokhov:2023ruj,Al-Badawi:2023lke,Guo:2024jhg,Zhang:2024nny,Pedrotti:2024znu,Gingrich:2024tuf,Bolokhov:2024voa,Skvortsova:2024eqi,Bonanno:2025dry,Skvortsova:2025cah,Konoplya:2025ect,Bolokhov:2025fto,Bolokhov:2025egl,Bolokhov:2025lnt,Lutfuoglu:2025pzi,Calza:2024fzo,Calza:2025yfm,Moura:2024vhz}.

Higher-curvature extensions of General Relativity are particularly well suited for such investigations. Among them, quasi-topological gravity occupies a distinguished position, as it allows for nontrivial higher-order curvature corrections while retaining second-order field equations for static, spherically symmetric spacetimes in arbitrary dimensions \cite{Boulware:1985wk,Wiltshire:1988uq,Cai:2003gr,Capozziello:2024ucm}. When combined with regular black hole metrics, quasi-topological gravity provides a consistent and tractable framework in which higher-curvature effects and regular cores jointly influence black hole radiation and dynamical properties \cite{Bueno:2024dgm,Konoplya:2024hfg,Konoplya:2024kih,Bueno:2024eig,Bueno:2025tli,Frolov:2026rcm}. Notice, that black holes in theories with usual Gauss-Bonnet or Lovelock corrections are usually singular and their quasinormal spectrum has been extensively studied (see, for instance \cite{Abdalla:2005hu,Moura:2006pz,Daghigh:2006xg,Zhidenko:2008fp,Yoshida:2015vua,Cuyubamba:2016cug,Konoplya:2017ymp,Konoplya:2017zwo,Kokkotas:2017zwt,Konoplya:2017lhs,Gonzalez:2018xrq,Zinhailo:2019rwd,Carson:2020ter,Konoplya:2020bxa,Cano:2020cao,Blazquez-Salcedo:2020caw,Konoplya:2020jgt,Blazquez-Salcedo:2022omw,Pierini:2022eim,Konoplya:2022iyn} and references therein).

In this work, we investigate grey-body factors and energy emission rates for higher-dimensional regular black holes arising in quasi-topological gravity. By solving the wave equations for the Maxwell field propagating in these backgrounds, we analyze how spacetime dimensionality and regularization parameter affect transmission probabilities and Hawking spectra. Special attention is devoted to deviations from purely thermal emission and to the role of the effective potential barrier surrounding the black hole. While Hawking radiation for two specific regular black hole solutions in quasi-topological gravity was previously considered in Ref.~\cite{Konoplya:2025uta} for particular values of the coupling parameter, a comprehensive analysis covering a broader class of regular solutions and the full range of parameters has not yet been performed.

The present work aims to fill this gap through a systematic study of grey-body factors and energy emission rates for multiple regular black hole configurations in quasi-topological gravity across several spacetime dimensions. By analyzing six distinct regular models and varying the coupling parameter up to its extremal limit, we identify common features and robust trends in the transmission coefficients and Hawking spectra. This approach allows us
to disentangle model-dependent quantitative differences from universal qualitative behavior, in particular the systematic suppression of emission and the emergence of a remnant-like extremal configuration. In this sense, the novelty of the present analysis lies not merely in extending known calculations to a new background, but in providing a unified and comparative framework for understanding semiclassical radiation from regular black holes in quasi-topological gravity.

In addition, we explore the correspondence between quasinormal modes and grey-body factors, treating it as a complementary aspect rather than a primary assumption. Quasinormal modes govern the characteristic damped oscillations of black holes and are closely related to the poles of scattering amplitudes in the complex frequency plane. It has been suggested in various contexts that features of the grey-body spectrum, such as characteristic frequency scales or transmission resonances, may be linked to the quasinormal mode spectrum \cite{Malik:2025dxn,Han:2025cal,Lutfuoglu:2025ldc,Bolokhov:2024otn,Dubinsky:2024vbn,Malik:2024cgb,Bolokhov:2025lnt,Dubinsky:2025nxv,Lutfuoglu:2025hjy,Malik:2025erb,Lutfuoglu:2025blw}. In the present work, we test this correspondence explicitly for higher-dimensional regular black holes in quasi-topological gravity by comparing the behavior of grey-body factors with the analytic approximation based on the quasinormal frequencies obtained for the same backgrounds.

This paper is designed as follows. In Sec.~\ref{sec:basic}, we introduce the regular black hole configurations in quasi-topological gravity and the master equations governing electromagnetic perturbations. In Sec.~\ref{sec:WKB}, we briefly review the WKB approach used to compute the grey-body factors. In Sec.~\ref{sec:energy}, we use the numerically obtained grey-body factors to evaluate the energy-emission rates and discuss the evaporation dynamics. Finally, in Sec.~\ref{sec:conclusions}, we summarize our results and discuss their physical implications.

\section{Maxwell field in the background of the quasi-topological black holes}\label{sec:basic}

Following Ref.~\cite{Bueno:2019ycr}, we consider a higher-curvature gravitational theory whose dynamics are governed by the action
\begin{equation}
    I_{QT} = \frac{1}{16\pi G}\int{d^D x\sqrt{|g|}\left[R+\sum^{n_{max}}_{n=2}{\alpha_n \mathcal{Z}_n}\right]},
    \label{eq:action}
\end{equation}
where \(G\) is the \(D\)-dimensional Newton constant and the quantities \(\mathcal{Z}_n\) denote quasi-topological curvature invariants of order \(n\). These terms are constructed so that, for static and spherically symmetric configurations, the resulting field equations reduce to second order, despite the presence of higher powers of the curvature. 

The static, spherically symmetric solutions of this theory can be written in the form \cite{Bueno:2024dgm}:
\begin{equation}
    ds^2 = -N(r)^2 f(r)dt^2 + \frac{dr^2}{f(r)} + r^2d\Omega^2_{D-2},
    \label{eq:metric}
\end{equation}
where \(\Omega^2_{D-2}\) denotes the line element of the unit 
\((D-2)\)-sphere, and, without loss of generality, we choose
\begin{equation}\label{eq:f(r)}
    N(r)=1,\qquad f(r)=1-r^2\psi(r).
\end{equation}
From the equations of motion it follows that the function \(\psi(r)\) satisfies the algebraic equation
\begin{equation}\label{eq:hpsi}
   h(\psi) \equiv \psi + \sum_{n=2}^{n_{max}}{\alpha_{n}\psi^{n}},\quad h(\psi)\equiv\frac{\mu}{r^{D-1}},
\end{equation}
where \(\mu\) is an integration constant that is directly related to the ADM mass of the black hole \(M\),
\begin{equation}\label{ADMmass}
    \mu\equiv\frac{16\pi G_{D} M}{(D-2)\Omega_{D-2}}, \quad \Omega_{D-2}=\frac{(2\pi)^{\frac{D-1}{2}}}{\Gamma(\frac{D-1}{2})}.
\end{equation}
For appropriate choices of the couplings \(\alpha_{n}\), this framework admits regular black hole solutions in which curvature invariants remain finite everywhere.
\begin{table}[t]
\centering
\small
\setlength{\tabcolsep}{4pt}
\renewcommand{\arraystretch}{1.25}
\begin{tabular*}{\textwidth}{@{\extracolsep{\fill}} c c c c}
\hline
Label 
& $f(r)$ 
& \makecell[c]{$\mu$ parameter in\\horizon units} 
& \makecell[c]{Constraint \\ in $\alpha$} \\
\hline

$(a)$ 
& $\displaystyle 1-\frac{\mu r^2}{r^{D-1}+\alpha \mu}$ 
& $\displaystyle \frac{r_0^{D-1}}{r_0^2-\alpha}$ 
& $\displaystyle 0 \le \frac{\alpha}{r_0^2} \le \frac{D-3}{D-1}$ \\

$(b)$ 
& $\displaystyle 1-\frac{\mu r^2}{\sqrt{r^{2(D-1)}+\alpha^2 \mu^2}}$ 
& $\displaystyle \frac{r_0^{D-1}}{\sqrt{r_0^4-\alpha^2}}$ 
& $\displaystyle 0 \le \frac{\alpha}{r_0^2} \le \sqrt{\frac{D-3}{D-1}}$ \\

$(c)$ 
& $\displaystyle 1-\frac{r^2}{\alpha}\left(1-e^{-\alpha \mu/r^{D-1}}\right)$ 
& $\displaystyle \frac{r_0^{D-1}}{\alpha}\ln\!\left(\frac{r_0^2}{r_0^2-\alpha}\right)$ 
& $\displaystyle 0 \le \frac{\alpha}{r_0^2}\leq C(D)<1$ \\

$(d)$ 
& $\displaystyle 1-\frac{2 \mu r^2}{r^{D-1}+\sqrt{r^{2(D-1)}+4 \alpha^2 \mu^2}}$ 
& $\displaystyle \frac{r_0^{D+1}}{r_0^4-\alpha^2}$ 
& $\displaystyle 0 \le \frac{\alpha}{r_0^2} \le \sqrt{\frac{D-3}{D+1}}$ \\

$(e)$ 
& $\displaystyle 1-\frac{2 \mu r^2}{r^{D-1}+2 \alpha \mu + \sqrt{r^{2(D-1)}+4 \mu \alpha r^{D-1}}}$ 
& $\displaystyle \frac{r_0^{D+1}}{(r_0^2-\alpha)^2}$ 
& $\displaystyle 0 \le \frac{\alpha}{r_0^2} \le \frac{D-3}{D+1}$ \\

$(f)$ 
& $\displaystyle 1-\frac{2 \mu r^2}{\mu \alpha + \sqrt{4 r^{2(D-1)}+\mu^2 \alpha^2}}$ 
& $\displaystyle \frac{r_0^{D-2}}{\sqrt{r_0^2-\alpha}}$ 
& $\displaystyle 0 \le \frac{\alpha}{r_0^2} \le \frac{D-3}{D-2}$ \\

\hline
\end{tabular*}
\caption{Summary of the considered regular black hole models. Configurations $(a)$–$(e)$ were proposed in \cite{Bueno:2024dgm}, while configuration $(f)$ was introduced in \cite{Arbelaez:2025gwj}. The constraint on $\alpha$ for configuration $(c)$, denoted $C(D)$, does not admit a closed-form expression. Its value increases monotonically with $D$, $C(5)\approx0.715$, $C(6)\approx0.802$, and approaches unity asymptotically, $\lim_{D\to\infty}C(D)=1$.}
\label{tab:solutions}
\end{table}

Tables~\ref{tab:solutions} and~\ref{tab:temperatures} summarize the different regular black hole geometries arising from infinite--curvature corrections that will be considered in this work. These spacetimes serve as background geometries for the computation of grey-body factors and the associated Hawking energy emission spectra.

\begin{table}[t]
\centering
\small
\setlength{\tabcolsep}{6pt}
\renewcommand{\arraystretch}{1.25}
\begin{tabular}{c c c}
\hline
Label & $h(\psi)$ & $4\pi T_H$ \\
\hline

$(a)$ 
&
$\displaystyle \frac{\psi}{1-\alpha\psi}$
& $\displaystyle \frac{1}{r_{0}^{3}}\left[(D-3)r_{0}^{2}-(D-1)\alpha\right]$ \\

$(b)$
&
$\displaystyle \frac{\psi}{\sqrt{1-\alpha^{2}\psi^{2}}}$
& $\displaystyle \frac{1}{r_{0}^{5}}\left[(D-3)r_{0}^{4}-(D-1)\alpha^{2}\right]$ \\

$(c)$
&
$\displaystyle -\frac{\log{\left(1-\alpha\psi\right)}}{\alpha}$
& $\displaystyle \frac{1}{r_{0}\alpha}
\left[(D-1)(r_{0}^{2}-\alpha)
\ln\!\left|\frac{r_{0}^{2}}{r_{0}^{2}-\alpha}\right|
-2\alpha\right]$ \\

$(d)$
&
$\displaystyle \frac{\psi}{1-\alpha^{2}\psi^{2}}$
& $\displaystyle \frac{(D-3)r_{0}^{4}-(D+1)\alpha^{2}}
{r_{0}(r_{0}^{4}+\alpha^{2})}$ \\

$(e)$
&
$\displaystyle \frac{\psi}{\left(1-\alpha\psi\right)^{2}}$
& $\displaystyle \frac{(D-3)r_{0}^{2}-(D+1)\alpha}
{r_{0}(r_{0}^{2}+\alpha)}$ \\

$(f)$
&
$\displaystyle \frac{\psi}{\sqrt{1-\alpha\psi}}$
& $\displaystyle \frac{2}{r_{0}}
\frac{(D-3)r_{0}^{2}-(D-2)\alpha}
{2r_{0}^{2}-\alpha}$ \\

\hline
\end{tabular}
\caption{Characteristic functions $h(\psi)$ and Hawking temperatures for the regular black hole models considered in this work.}
\label{tab:temperatures}
\end{table}

Before proceeding, it is important to comment on the physical viability and stability of these background solutions. A complete analysis of gravitational perturbations in quasi-topological gravity is technically involved and lies beyond the scope of the present paper. In particular, establishing the absence of ghost or gradient instabilities requires studying the linearized field equations of the underlying higher-curvature theory and verifying that no additional propagating degrees of freedom exhibit pathological behavior. Here, we restrict ourselves to parameter ranges for which the solutions are regular, possess a well-defined event horizon, and satisfy the standard consistency requirements of quasi-topological gravity, thereby avoiding known pathological branches of the theory. Moreover, the dynamical formation of regular black holes in related higher-curvature frameworks has recently been demonstrated, providing further support for the physical relevance of such geometries \cite{Bueno:2024eig}.

At the level relevant for the present work, namely test-field perturbations propagating on these fixed backgrounds, there was found no indication of dynamical instabilities. The effective potentials governing electromagnetic perturbations remain positive-definite outside the event horizon for all models and admissible values of the coupling parameter $\alpha$, preventing the appearance of exponentially growing modes. The associated quasinormal-mode spectra exhibit the expected damped behavior, with negative imaginary parts, supporting linear stability in the probe-field sector \cite{Konoplya:2024hfg,Dong:2024ams,Arbelaez:2025gwj}. Although a full gravitational perturbation analysis would be required to establish complete dynamical stability, the absence of instabilities in the test-field sector ensures that the geometries considered here provide a consistent framework for investigating grey-body factors and semiclassical radiation processes.

In the present work, we study the dynamics of test fields propagating on the background of these regular black hole spacetimes. In particular, we focus on electromagnetic perturbations described by the Maxwell equations in \(D\)-dimensional curved spacetime,
\begin{align}
\label{Maxwell}
\frac{1}{\sqrt{-g}} \partial_\mu \Big( \sqrt{-g} \, F_{\rho\sigma} g^{\rho\nu} g^{\sigma\mu} \Big) &= 0,
\end{align}
where 
\(F_{\mu\nu} = \partial_\mu A_\nu - \partial_\nu A_\mu\) is the electromagnetic field strength tensor associated with the vector potential \(A_\mu\).

Following Ref.~\cite{Crispino:2000jx}, we adopt Feynman’s gauge, which allows a separation of variables. After decomposing the vector potential into scalar-type and vector-type components on the \((D-2)\)-sphere, the perturbation equations reduce to a set of one–dimensional wave equations of Schrödinger type,
\begin{equation}\label{eq:wavelike}
    \frac{d^2 \Psi(r_*)}{dr_*^2} + \left[ \omega^2 - V(r_*) \right] \Psi(r_*) = 0,
\end{equation}
where \(\omega\) is the frequency of the perturbation and where \(r_*\) is the tortoise coordinate defined by 
\[dr_* \equiv \frac{dr}{f(r)}.\]

The effective potentials governing the scalar-type and vector-type electromagnetic perturbations are given, respectively, by \(V_S(r)\) and \(V_V(r)\) \cite{Lopez-Ortega:2006vjp}:
\begin{subequations}
\begin{align}
\label{eq:V1}
V_S(r) &= f(r) \left( 
\frac{\ell(\ell+D-3)}{r^2} 
+ \frac{(D-2)(D-4)}{4r^2}f(r) 
- \frac{D-4}{2r}\frac{df}{dr}
\right), \\
\label{eq:V2}
V_V(r) &= f(r) \left( 
\frac{(\ell+1)(\ell+D-4)}{r^2} 
+ \frac{(D-4)(D-6)}{4r^2}f(r) 
+ \frac{D-4}{2r}\frac{df}{dr}
\right),
\end{align}
\end{subequations}

where $\ell = 1,2,\dots$ denotes the multipole number.

\section{WKB approach for the calculation of grey-body factors}\label{sec:WKB}
We consider the scattering problem associated with Eq.~(\ref{eq:wavelike}), subject to the boundary conditions
\begin{subequations}
\begin{align}
    \Psi &= e^{-i\omega r_{*}} + R\, e^{i\omega r_{*}}, && r_{*} \to +\infty, \label{eq:psi_infinity} \\
    \Psi &= T\, e^{-i\omega r_{*}}, && r_{*} \to -\infty. \label{eq:psi_horizon}
\end{align}
\end{subequations}
which describe, respectively, an incident wave from spatial infinity partially reflected by the potential barrier and a purely ingoing wave at the event horizon. The coefficients \(T\) and \(R\) correspond to the transmission and reflection amplitudes, and their moduli squared correspond to the transmission and reflection coefficients. Then, for a given real frequency \(\omega\), the grey-body factor is defined as
\begin{equation}\label{greybody}
    \Gamma_{\ell}(\omega)\equiv|T|^2=1-|R|^{2}.
\end{equation}

Applying the WKB approximation to Eq.~(\ref{eq:wavelike}) allows one to compute the scattering coefficients in terms of the WKB phase through the relation
\begin{equation}
   \Gamma_{\ell}(\omega) = \frac{1}{1+e^{2i\pi\mathcal{K}}},
\end{equation}
where $\mathcal{K}$ is a function of the frequency \(\omega\) determined by the properties of the effective potential near its maximum. 

Following the original formulation of Schutz and Will \cite{Schutz:1985km}, we employ the WKB expansion corrected to higher orders derived in \cite{Iyer:1986np,Konoplya:2003ii,Matyjasek:2017psv,Matyjasek:2019eeu}. The resulting expression reads
\begin{equation}
    \mathcal{K} \;=\; i\frac{\omega^{2}-V_0}{\sqrt{-2V_0''}} - \sum_{k=2}^{N} \Lambda_{k}.
    \label{eq:correctedomega}
\end{equation}
where $V_0$ denotes the value of the effective potential at its maximum \(r=r_{max}\), and $V_0''$ is the second derivative of the potential with respect to the tortoise coordinate evaluated at the same point. The correction terms $\Lambda_{k}$ depend on higher-order derivatives of the potential at \(r_{max}\). The above WKB approach was broadly used for finding quasinormal modes and grey-body factors of black holes \cite{Konoplya:2005sy,Konoplya:2019xmn,MahdavianYekta:2019pol,Konoplya:2020cbv,Matyjasek:2021xfg,Konoplya:2021ube,Konoplya:2023moy,Bolokhov:2023bwm,Dubinsky:2024nzo,Bolokhov:2024ixe,Skvortsova:2024wly,Malik:2024nhy,Miyachi:2025ptm,Hamil:2025cms,Pedrotti:2025idg,Lutfuoglu:2025ljm,Antonelli:2025yol,Zhang:2025xqt}.

The WKB approximation yields reliable transmission coefficients only within an intermediate frequency regime. At low frequencies, 
\(\omega^{2}\ll V_{\mathrm{0}}\), the classical turning points become widely separated, and the potential can no longer be accurately approximated by a local expansion around its maximum. Consequently, the WKB series fails already at leading order, and the inclusion of higher-order corrections does not improve the accuracy but instead amplifies deviations from the exact solution. In the opposite high-frequency regime, 
\(\omega^{2}\gg V_{\mathrm{0}}\), the potential barrier becomes effectively transparent and the scattering coefficients approach their geometric-optics limit. In this case, the WKB expansion loses its asymptotic character, and higher-order terms introduce numerical instabilities rather than enhancing the precision of the approximation. These limitations are intrinsic to the WKB method and cannot be remedied by simply increasing the order of the expansion.

\begin{figure}[t]
\centering

\begin{subfigure}{0.495\textwidth}
    \centering
    \includegraphics[width=\linewidth]{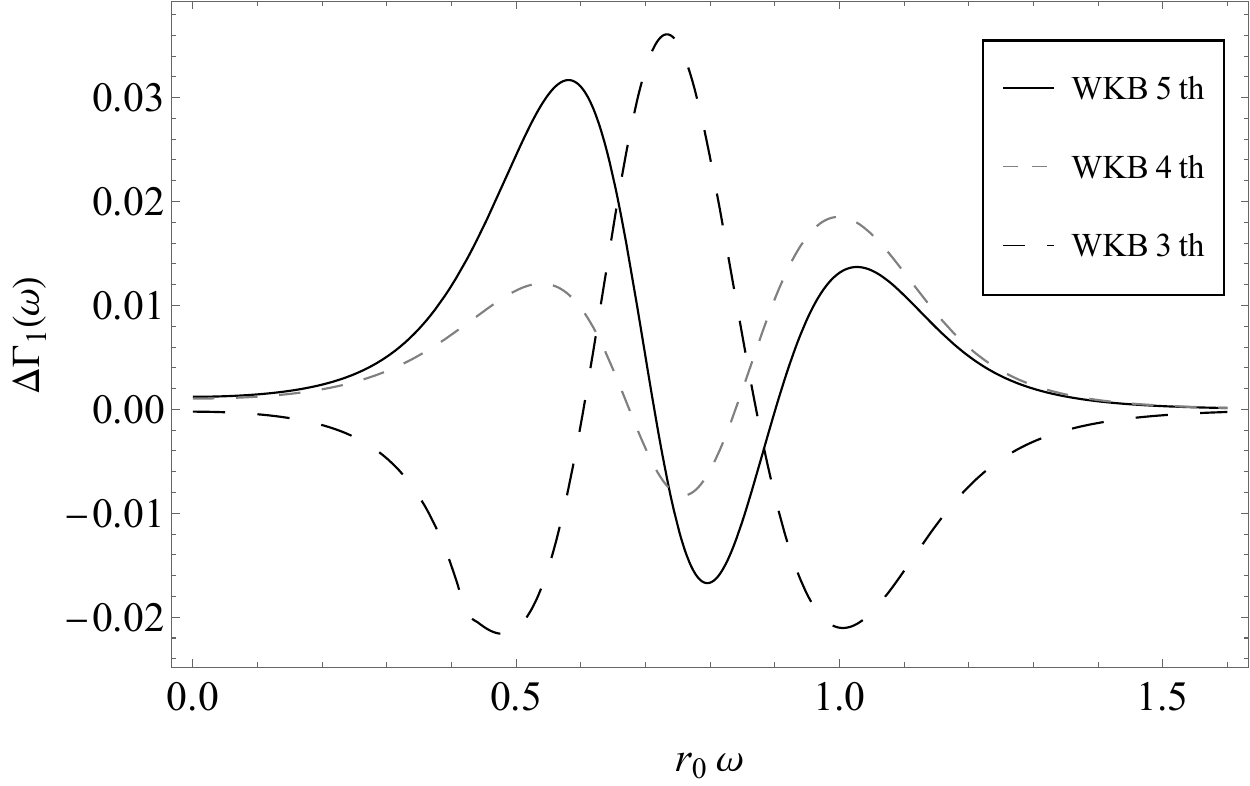}
\end{subfigure}
\hfill
\begin{subfigure}{0.495\textwidth}
    \centering
    \includegraphics[width=\linewidth]{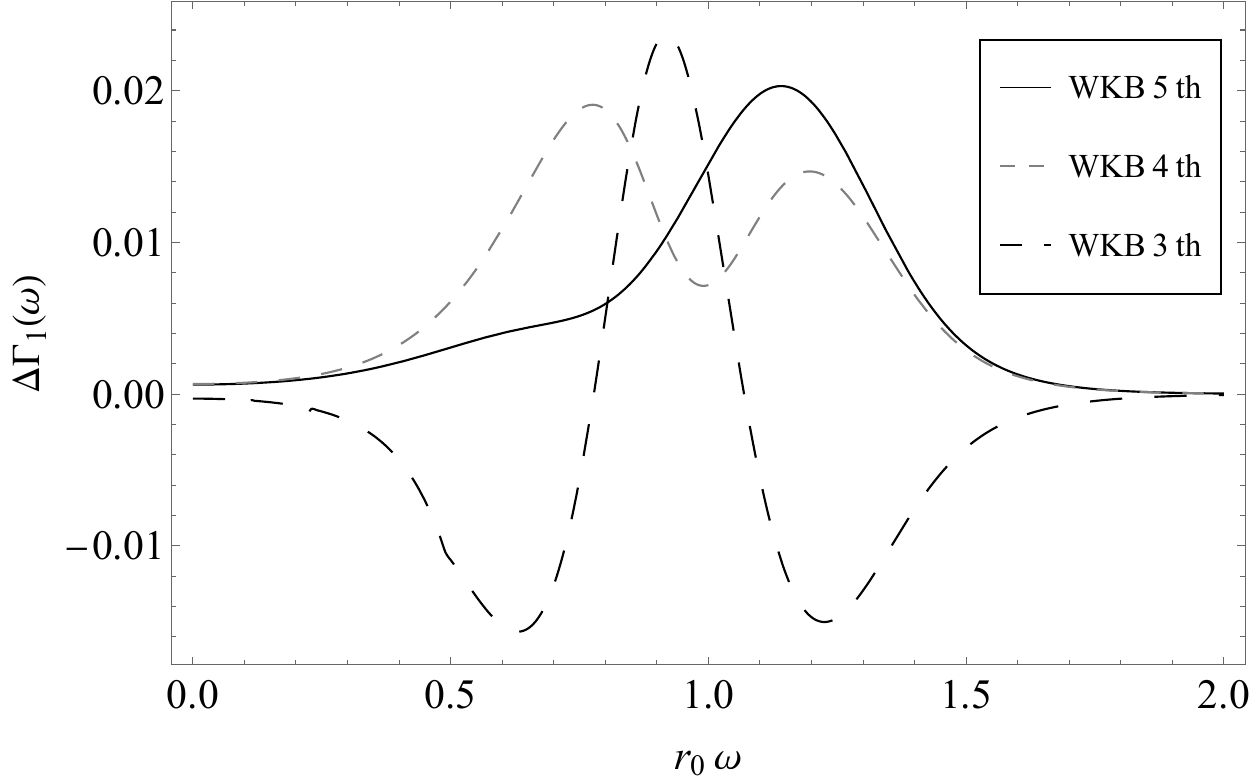}
\end{subfigure}

\caption{
Difference between WKB approximations of various orders and the accurate grey-body data for electromagnetic perturbations 
of the $D=5$ black hole $(f)$ for $\ell=1$ and $\alpha=0.3r_0^2$: 
$V_S$~(\ref{eq:V1}) (left) and $V_V$~(\ref{eq:V2}) (right). Accurate data are taken from ref.~\cite{Konoplya:2025uta}.}
\label{fig:diff}
\end{figure}

Taking into account both the loss of asymptotic reliability at very high WKB orders and the insufficient accuracy of the lowest orders, we restrict our analysis to intermediate orders of the expansion. A direct comparison between different truncation orders, supported by the numerical and graphical evidence presented in Fig.~\ref{fig:diff}, indicates that the fourth- ($N=4$) and fifth-order ($N=5$) WKB approximations provide the most stable and accurate results over the frequency range relevant to our analysis. In particular, the relative deviation between the WKB transmission coefficients and the accurate numerical data remains within several percent throughout the dominant frequency interval contributing to the Hawking flux. This quantitative uncertainty is significantly smaller than the physical effect displayed in Fig.~\ref{fig:reflecoeff}, where the suppression of the transmission coefficients due to increasing $\alpha$ leads to variations that are substantially larger than the estimated WKB error. Therefore, while the approximation has well-defined limitations outside the intermediate regime, the observed suppression of grey-body factors cannot be attributed to truncation artifacts of the WKB method, but represents a genuine physical feature of the regular black hole geometries considered.

An alternative method for estimating the grey-body factors was proposed in Ref.~\cite{Konoplya:2024lir,Konoplya:2024vuj}. This approach expresses the transmission coefficient in terms of the two dominant quasinormal modes, 
\(\omega_{0}\) and \(\omega_{1}\). The quasinormal mode spectrum for the present class of spacetimes was studied using the WKB method in Ref.~\cite{Arbelaez:2025gwj}. The resulting approximate expression for the grey-body factor reads
\begin{equation}
    \Gamma_{\ell}(\omega) \approx \left[ 1 + \exp\left( \frac{2\pi\bigl(\omega^{2}-\mathrm{Re}(\omega_{0})^{2}\bigr)}{4\,\mathrm{Re}(\omega_{0})\,\mathrm{Im}(\omega_{0})} \right) \right]^{-1}+\mathcal{O}(\ell^{-1}),
    \label{eq:approx}
\end{equation}
which provides accurate results for low multipole numbers and becomes exact in the eikonal limit $\ell\to\infty$; its accuracy is further enhanced by the inclusion of higher-order corrections arising from overtone contributions. The above correspondence has been applied and tested in a number of recent publications \cite{Skvortsova:2024msa,Lutfuoglu:2025ohb,Lutfuoglu:2025kqp,Lutfuoglu:2025mqa,Huang:2025rxx,Liang:2026eic}, showing as a rule a good concordance for higher multipole numbers, provided the effective potential has a single peak.  For double well potentials the correspondence is usually broken \cite{Konoplya:2025hgp}. Notice that the grey-body factors are usually more stable to small deformations of the background than quasinormal modes \cite{Oshita:2024fzf,Konoplya:2025ixm}. 

\begin{figure}[t]
\centering

\begin{subfigure}{0.495\textwidth}
    \centering
    \includegraphics[width=\linewidth]{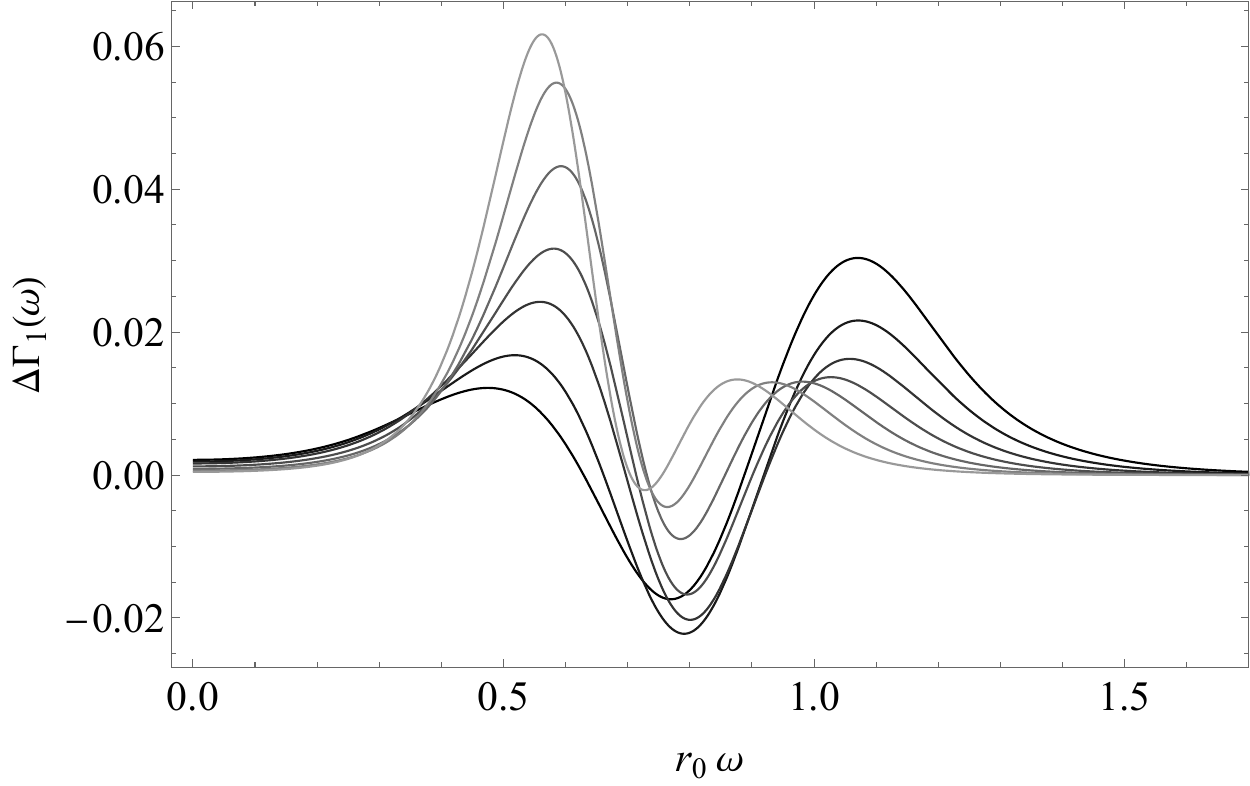}
\end{subfigure}
\hfill
\begin{subfigure}{0.495\textwidth}
    \centering
    \includegraphics[width=\linewidth]{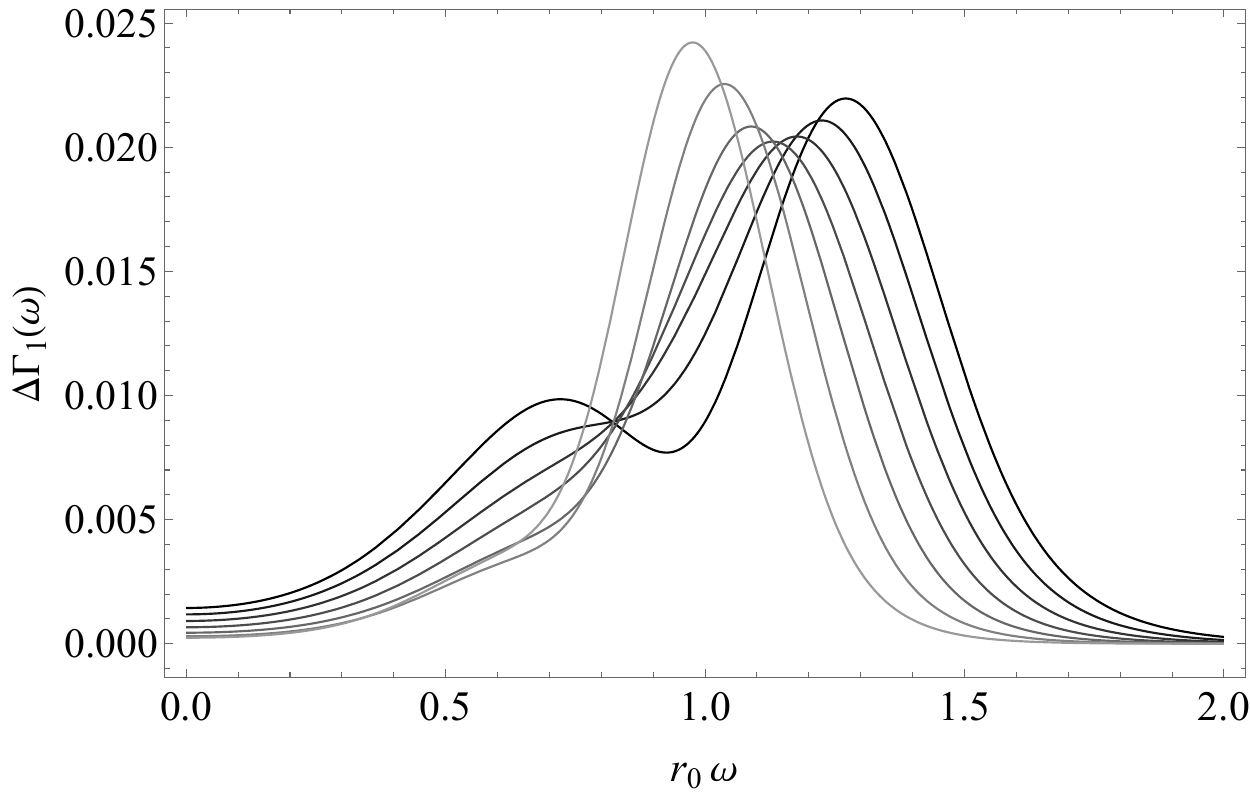}
\end{subfigure}

\caption{Difference between the WKB approximation and the analytic expression~\eqref{eq:approx}, expanded up to order $\mathcal{O}(\ell^{-3})$, for electromagnetic perturbations of the $D=5$ black-hole configuration $(f)$ with $\ell=1$. The curves correspond to different values of the coupling parameter $\alpha$ ($0, 0.1, \dots, 0.6$), with lighter curves indicating larger $\alpha$. The left panel shows the scalar-type potential $V_S$~(\ref{eq:V1}), while the right panel shows the vector-type potential $V_V$~(\ref{eq:V2}).}
\label{fig:diff2}
\end{figure}

To assess the validity of the quasinormal-mode/grey-body correspondence, in Fig.~\ref{fig:diff2} we explicitly compare the grey-body factors obtained within the WKB approximation with those derived from the expression (3.5), finding good agreement within several percent over the relevant frequency range.

\section{Energy emission rate}\label{sec:energy}

\begin{figure}[t]
    \begin{subfigure}{0.5\textwidth}
        \includegraphics[width=\linewidth]{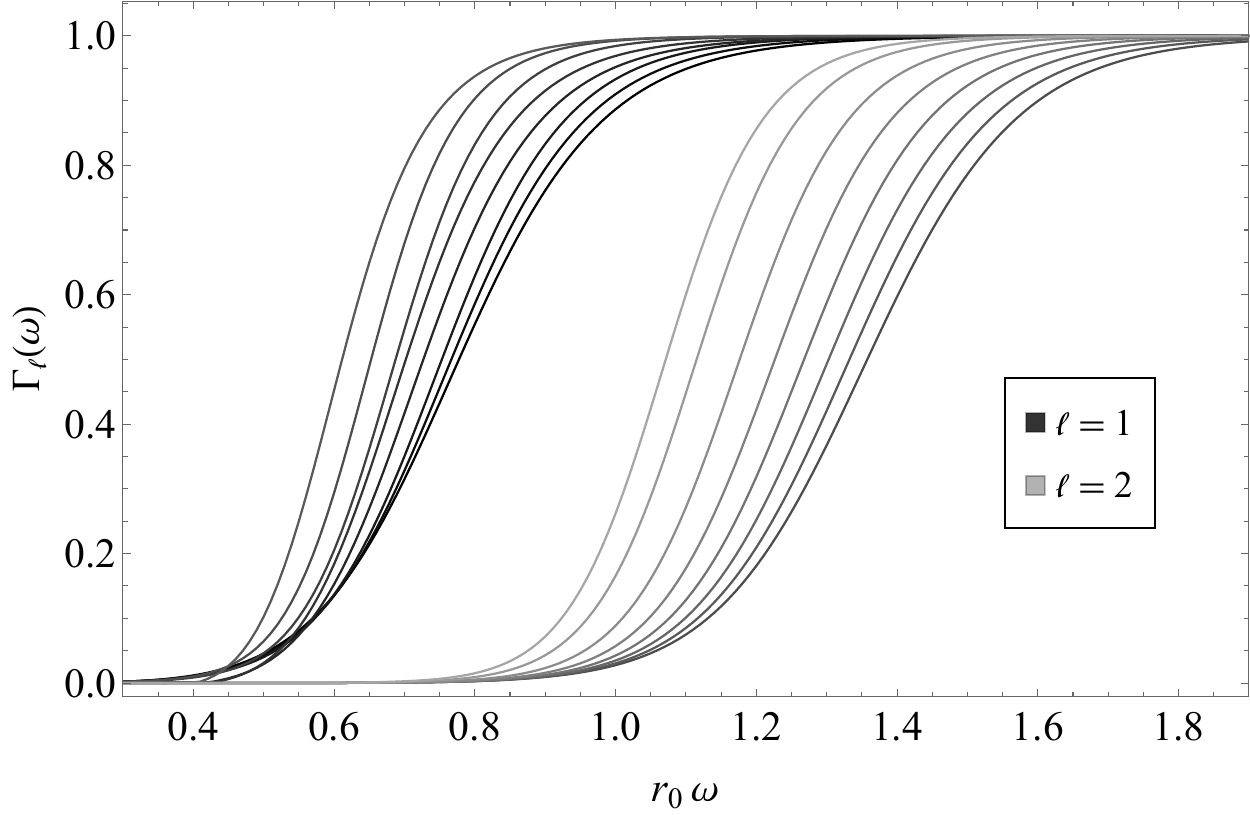}
    \end{subfigure}
    \hfill
    \begin{subfigure}{0.5\textwidth}
        \includegraphics[width=\linewidth]{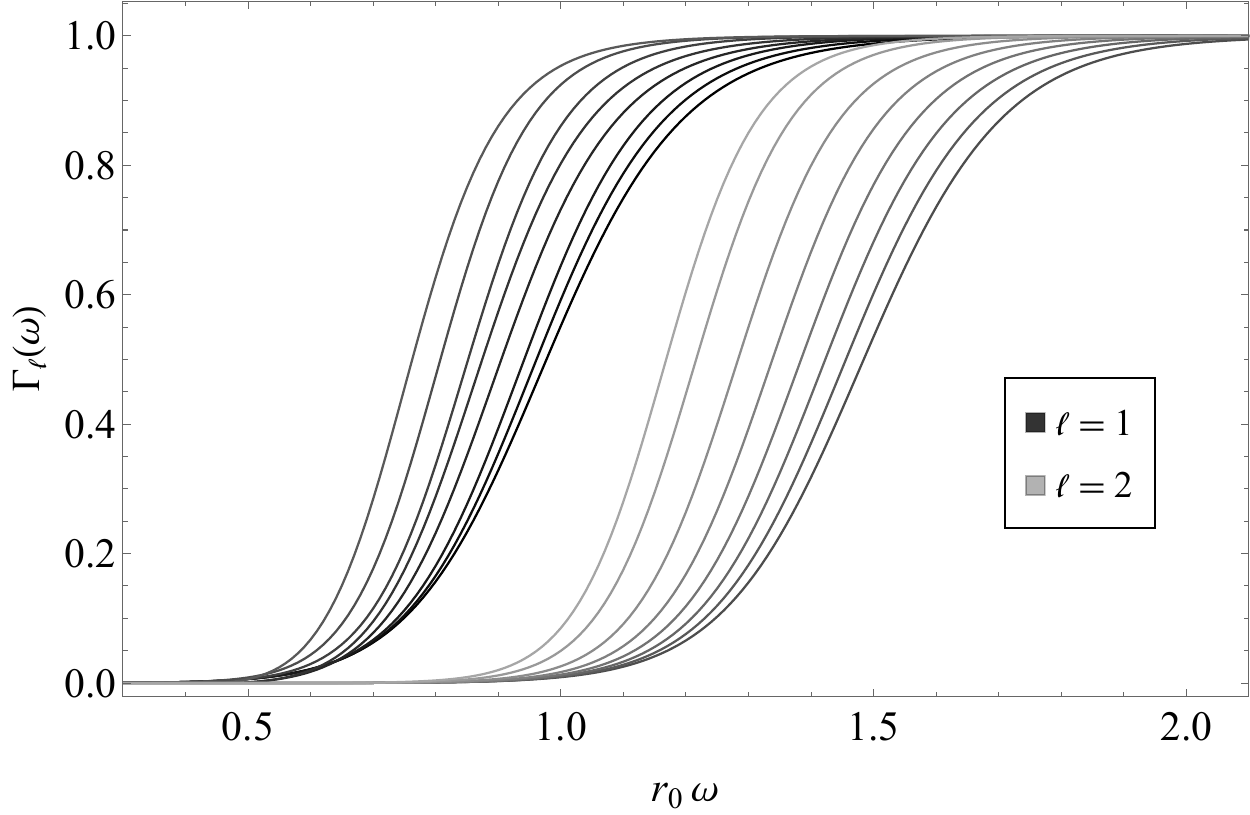}
    \end{subfigure}
    \caption{Transmission coefficients for electromagnetic perturbations 
    of the \(D=5\) black hole $(f)$: $V_{S}$~(\ref{eq:V1}) (left) and $V_{V}$~(\ref{eq:V2}) (right). 
    The curves show the dependence on the parameter 
    \(\alpha = 0, 0.1, \ldots, 0.6, 0.66\), increasing from darker to lighter tones.}
    \label{fig:reflecoeff}
\end{figure}

In Fig.~\ref{fig:reflecoeff}, we display the grey-body factors for the black-hole model \((f)\). For small values of the coupling parameter \(\alpha\), transmission occurs only at relatively high frequencies. As \(\alpha\) increases, the effective potential barrier becomes higher and broader, thereby reducing the frequency range for which wave propagation is allowed. Consequently, low-frequency modes are increasingly suppressed with growing \(\alpha\). This qualitative behavior is consistently observed for all remaining configurations considered in this work.

Using the numerically computed grey-body factors, one can evaluate the energy-emission rate via
\begin{equation}\label{EErate}
    \frac{\partial^{2} E}{\partial \omega \partial t}=\frac{1}{2\pi}\sum_{\ell}{N_{\ell}\Gamma_{\ell}(\omega)\frac{\omega}{\exp{(\omega/T_{H})-1}}},
\end{equation}
where $T_H$ denotes the Hawking temperature (listed in Table~\ref{tab:temperatures}),
\begin{equation}\label{eq:Hawking}
T_H = \frac{f'(r_{0})}{4\pi},    
\end{equation}
and \(N_{\ell}\) are the multiplicity factors accounting for the degeneracy of electromagnetic modes in \(D\)-dimensional spacetime. For scalar-type (\(V_S\)) and vector-type (\(V_V\)) perturbations, these factors are given by \cite{Kanti:2004nr}
\begin{eqnarray}
    N_{\ell}^{(S)}&=&\frac{(2\ell+D-3)(\ell+D-4)!}{\ell!(D-3)!},\\\nonumber
    N_{\ell}^{(V)}&=&\frac{(\ell+D-3)\ell(2\ell+D-3)(\ell+D-5)!}{(\ell+1)!(D-4)!}.
\end{eqnarray}
The above multiplicity take into account the degeneracy of the electromagnetic modes in $D$-dimensional spacetime.

\begin{figure}[t]
\centering
    \begin{subfigure}{0.495\textwidth}
        \centering
        \includegraphics[width=\linewidth]{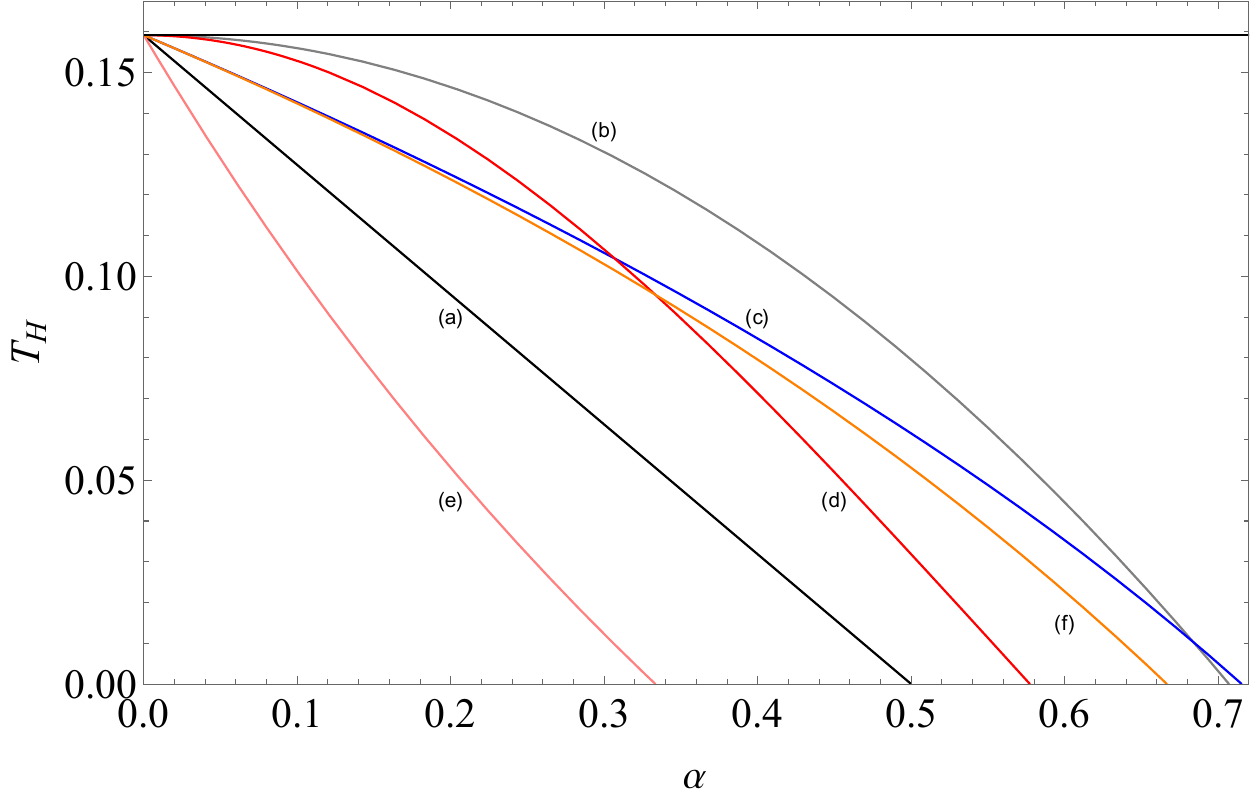}
    \end{subfigure}
    \hfill
    \begin{subfigure}{0.495\textwidth}
        \centering
        \includegraphics[width=\linewidth]{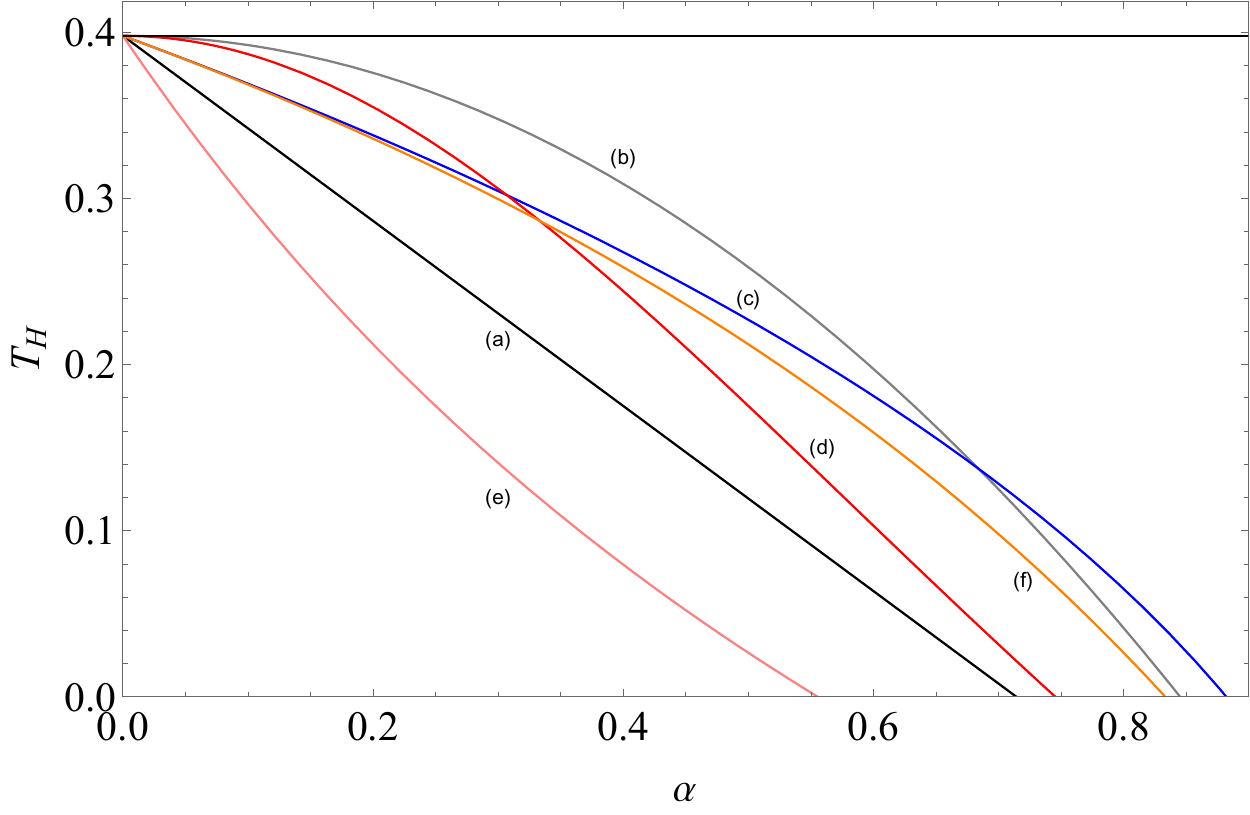}
    \end{subfigure}
\caption{Hawking temperature for $D=5$ and $D=8$ , respectively, with $r_{0}=1$ for various regular black holes. The upper solid black curve corresponds to the Tangherlini solution. The remaining curves represent the regular configurations: black line $(a)$, grey line $(b)$, blue line $(c)$, red line $(d)$, pink line $(e)$, and orange line $(f)$.}
\label{fig:TH}
\end{figure}

Figure~\ref{fig:TH} displays the Hawking temperature for the configurations listed in Table~\ref{tab:solutions} as a function of the coupling parameter $\alpha$. The plot illustrates how the higher-curvature corrections affect the thermodynamic behaviour of the black hole, leading in particular to a monotonic decrease of $T_H$ as the corrections become stronger.

In addition, even if the WKB approximation loses accuracy in the very low-frequency regime, its impact on the total energy emission rate remains negligible. This is because the infrared contribution is strongly regulated by the thermal Bose--Einstein factor $(\exp{(\omega/T_H)}-1)^{-1}$. For $\omega \ll T_H$, one has the expansion
\[
\frac{\omega}{\exp{(\omega/T_H)}-1}
=
T_H + \mathcal{O}(\omega^{2}),
\]
so that the emission spectrum scales as $\sim T_H\,\Gamma_\ell(\omega)$. Therefore, in the infrared regime the spectral behavior is entirely controlled by the grey-body factor. Since $\Gamma_\ell(\omega)\to 0$ as $\omega\to 0$ for $\ell\ge1$, the low-frequency contribution to the integrated flux is naturally suppressed. Consequently, possible inaccuracies of the WKB approximation in this regime do not significantly affect the total radiative power.

\begin{figure}[t]
\centering
    \begin{subfigure}{0.495\textwidth}
        \centering
        \includegraphics[width=\linewidth]{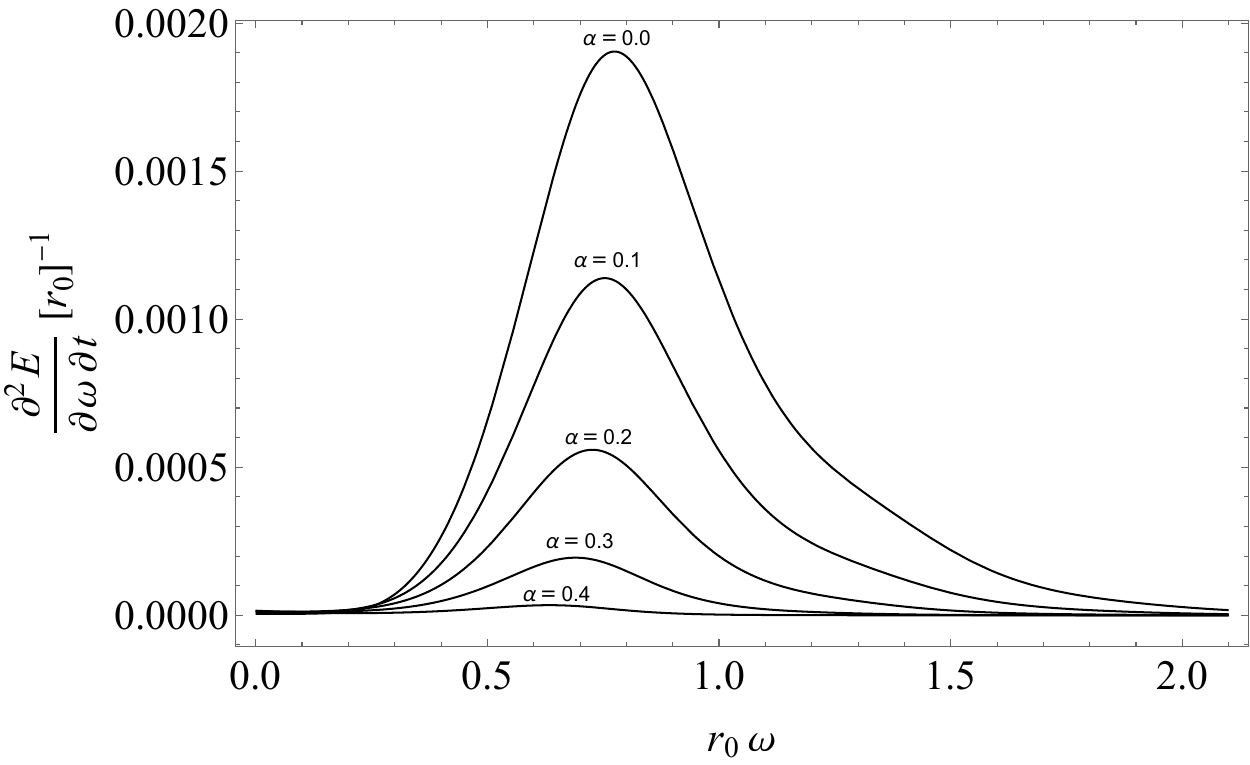}
    \end{subfigure}
    \hfill
    \begin{subfigure}{0.495\textwidth}
        \centering
        \includegraphics[width=\linewidth]{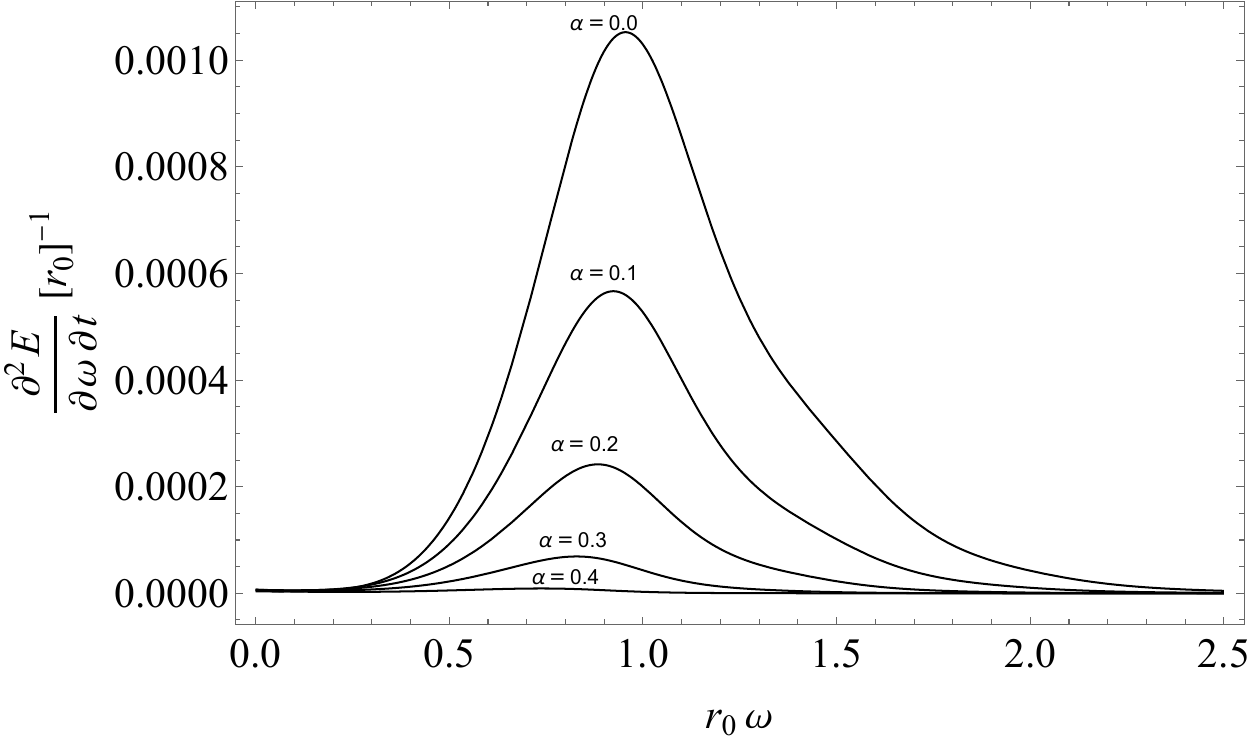}
    \end{subfigure}
\caption{Energy emission rates for the electromagnetic perturbations of the $D=5$ black hole $(f)$: $V_{S}$~(\ref{eq:V1}) (left) and $V_{V}$~(\ref{eq:V2}) (right), for different values of the parameter $\alpha$.}
\label{fig:energyf}
\end{figure}

In Fig.~\ref{fig:energyf}, we present the energy-emission rates for different values of \(\alpha/r_0^{2}\). Increasing \(\alpha\) lowers the Hawking temperature, shifting the peak of the emission spectrum toward lower frequencies and suppresses both the frequency range of significant emission and the total radiated power. In the $D=5$ case for configuration $(f)$, the multipolar decomposition shows that higher--$\ell$ modes decay extremely rapidly. The $\ell=1$ mode contributes more than $95\%$ of the total radiative flux, while already for $\ell=5$ the contribution drops to $\sim 1.8\times10^{-5}\%$ of the total emission. Hence, the contribution of the eikonal sector becomes negligibly small. Since the WKB approximation is asymptotically exact in the eikonal limit $\ell\to\infty$, and the suppression persists up to that regime, the effect cannot be attributed to a breakdown of the approximation scheme. Rather, it reflects an intrinsic property of the underlying geometry and its associated effective potential. A qualitatively similar behaviour is observed across the other models considered in this work.

By numerically integrating the emission spectra, we obtain the total energy radiated per unit time for each effective potential.

\begin{table}[t]
\centering
\begin{tabular}{c S S}
\hline
{$\alpha$} & {$\Xi_S$} & {$\Xi_V$} \\
\hline
0   & 1.13194e-3 & 7.14534e-4 \\
0.1 & 6.13596e-4 & 3.51646e-4 \\
0.2 & 2.72738e-4 & 1.37896e-4 \\
0.3 & 8.75558e-5 & 3.74379e-5 \\
0.4 & 1.59197e-5 & 5.51799e-6 \\
0.5 & 1.39108e-6 & 4.59524e-7 \\
0.6 & 1.26795e-7 & 4.34701e-8 \\
\hline
\end{tabular}
\caption{Energy emitted per unit time for the electromagnetic perturbations ($D=5$, $r_0=1$): $V_{S}$~(\ref{eq:V1}) and $V_{V}$~(\ref{eq:V2}) for different values of $\alpha$ for black hole $(f)$.}
    \label{tab:energy_V1V2}
\end{table}

The total energy-emission rate includes a sum over all perturbation types and channels,
\begin{equation}\label{energyemission}
\frac{dM}{dt}=-\frac{dE}{dt}
    \;\simeq\;
-\,\frac{1}{r_{0}^2}\sum_{i} \Xi_i(r_0),
\end{equation}
where \(\Xi_i(r_0)\) represents the individual contributions expressed in units set by the horizon radius \(r_0\). In the present paper we estimate only \(\Xi_S\) and \(\Xi_V\), corresponding to electromagnetic emission in the scalar- and vector-type channels, respectively. The results are summarized in Table~\ref{tab:energy_V1V2}: As anticipated from the behavior of the grey-body factors and the decreasing Hawking temperature, the total emission rate drops rapidly as $\alpha$ increases, spanning several orders of magnitude.

\begin{table}[ht]
    \centering
    \begin{tabular}{c c c c c c}
    \hline
    \text{\makecell{Model}} &
    \text{\makecell{Type of\\potential}} &
    $A$ & $a$ & $b$ & $c$ \\
    \hline
    \multirow{2}{*}{$(a)$} 
        & $V_S$ & $0.587262$ & $-0.342696$ & $0.790843$ & $-8.7833$ \\
        & $V_V$ & $0.391783$ & $-0.174249$ & $0.0688829$ & $-6.9123$ \\
    \hline
    \multirow{2}{*}{$(b)$} 
        & $V_S$ & $0.587396$ & $-0.00152796$ & $0.0369438$ & $-0.758508$ \\
        & $V_V$ & $0.391747$ & $-0.0161219$ & $0.341681$ & $-1.05769$ \\
    \hline
    \multirow{2}{*}{$(c)$} 
        & $V_S$ & $0.586869$ & $-0.116093$ & $-0.184914$ & $-0.826694$ \\
        & $V_V$ & $0.391408$ & $-0.0586019$ & $-0.105336$ & $-1.16548$ \\
    \hline
    \multirow{2}{*}{$(d)$} 
        & $V_S$ & $0.587262$ & $-0.0198551$ & $0.2358$ & $-2.01235$ \\
        & $V_V$ & $0.389003$ & $-0.0268256$ & $0.681175$ & $-2.7413$ \\
    \hline
    \multirow{2}{*}{$(e)$} 
        & $V_S$ & $0.587263$ & $-1.26255$ & $11.0673$ & $-53.7883$ \\
        & $V_V$ & $0.391783$ & $-0.144991$ & $-4.87879$ & $-11.3966$ \\
    \hline
    \multirow{2}{*}{$(f)$} 
        & $V_S$ & $0.606061$ & $-0.149416$ & $-0.211663$ & $-0.972222$ \\
        & $V_V$ & $0.382579$ & $-0.0136779$ & $-0.177707$ & $-1.10004$ \\
    \hline
    \end{tabular}
    \caption{Best–fit values of the parameters $A$, $a$, $b$, and $c$ extracted from the exponential approximation of the energy emission rate for the different black-hole models in $D=5$  ($r_{0}=1$).}
    \label{tab:interpD5}
\end{table}
\begin{table}[ht]
    \centering
    \begin{tabular}{c c c c c c}
    \hline
    \text{\makecell{Model}} &
    \text{\makecell{Type of\\potential}} &
    $A$ & $a$ & $b$ & $c$ \\
    \hline
    \multirow{2}{*}{$(a)$} 
        & $V_S$ & $0.321374$ & $0.400237$ & $-1.17956$ & $-0.73317$ \\
        & $V_V$ & $0.248383$ & $0.936455$ & $-0.363306$ & $-5.89591$ \\
    \hline
    \multirow{2}{*}{$(b)$} 
        & $V_S$ & $0.334118$ & $-0.0449428$ & $0.889182$ & $-1.90066$ \\
        & $V_V$ & $0.251735$ & $-0.0187384$ & $1.47931$ & $-1.2256$ \\
    \hline
    \multirow{2}{*}{$(c)$} 
        & $V_S$ & $0.321297$ & $0.192584$ & $-0.13182$ & $-0.57329$ \\
        & $V_V$ & $0.248185$ & $0.492866$ & $-0.0276431$ & $-1.0298$ \\
    \hline
    \multirow{2}{*}{$(d)$} 
        & $V_S$ & $0.321265$ & $-0.0901956$ & $2.18694$ & $-5.4443$ \\
        & $V_V$ & $0.248374$ & $-0.055867$ & $3.01892$ & $-5.15982$ \\
    \hline
    \multirow{2}{*}{$(e)$} 
        & $V_S$ & $0.321374$ & $0.521551$ & $-1.3333$ & $-9.19442$ \\
        & $V_V$ & $0.248383$ & $1.83655$ & $-4.45207$ & $-17.6569$ \\
    \hline
    \multirow{2}{*}{$(f)$} 
        & $V_S$ & $0.337076$ & $0.2576$ & $-0.113645$ & $-0.677156$ \\
        & $V_V$ & $0.248322$ & $0.498347$ & $0.03661$ & $-1.20533$ \\
    \hline
    \end{tabular}
    \caption{Best–fit values of the parameters $A$, $a$, $b$, and $c$ extracted from the exponential approximation of the energy emission rate for the different black-hole models in $D=6$  ($r_{0}=1$).}
    \label{tab:interpD6}
\end{table}

The strong suppression of the emission rate as \(\alpha\) approaches its upper bound suggests the exponential dependence governed primarily by the Hawking temperature given by Eq.~(\ref{eq:Hawking}). We fit the numerical data using the following function:
\begin{equation}\label{eq:Xi}
\displaystyle
\Xi_i(r_0) \;=\; A_i\,
\exp\!\left(
-\,\frac{1+a_i\dfrac{\alpha}{r_0^2}+b_i\dfrac{\alpha^2}{r_0^4}+c_i\dfrac{\alpha^3}{r_0^6}}
{r_{0}\,T_H(r_{0})}
\right),
\end{equation}
which provides an accurate description of both scalar-type and vector-type electromagnetic emission for all black-hole configurations considered. The numerical values of the coefficients \(a_i\), \(b_i\), \(c_i\), and \(A_i\) are reported in Tables~(\ref{tab:interpD5}-\ref{tab:interpD6}). As \(\alpha\) increases, the dominant contribution to the exponent arises from the inverse Hawking temperature \(1/T_{H}\), since \(T_{H}\) decreases more rapidly than the geometric corrections encoded in higher-order terms.

To express the evaporation dynamics in terms of the geometric size of the black hole, we use Eq.~(\ref{ADMmass}) to write the mass as a function of the horizon radius,
\begin{equation}\label{eq:Mr0}
M(r_{0})=\frac{(D-2)\Omega_{D-2}}{16\pi G_{D}}\mu(r_{0}),
\end{equation}
where, from Eq.~(\ref{eq:hpsi}),
\[\mu(r_0)=r_{0}^{D-1}h\left(\frac{1}{r_{0}^{2}}\right),\]
with explicit expressions for each model given in Table~\ref{tab:solutions}.

Differentiating Eq.~(\ref{eq:Mr0}) and using Eqs.~(\ref{eq:f(r)}), (\ref{eq:hpsi}), and (\ref{eq:Hawking}), after some algebra we find
\begin{equation}\nonumber
    \frac{dM}{dt}=\frac{dr_{0}}{dt}\frac{(D^{2}-3D+2)\Omega_{D-2}T_{H}(r_{0})}{8G_{D}(1+2\pi r_{0}T_{H}(r_{0}))}r_{0}^{D-1}h\left[\frac{1}{r_{0}^{2}}\right].
\end{equation}

Combining this expression with the law on energy-emissions~(\ref{energyemission}), we obtain the differential equation for the horizon radius,
\begin{equation}\label{eq:eed}
    \frac{dr_{0}}{dt}=\frac{8G_{D}(1+2\pi r_{0}T_{H}(r_{0}))\sum_{i} \Xi_i(r_0)}{(D^{2}-3D+2)\Omega_{D-2}T_{H}(r_{0})r_{0}^{D+1}h\left[\frac{1}{r_{0}^{2}}\right]}.
\end{equation}

Assuming that the energy-emission rates for all types of perturbation and channels exhibit the same qualitative dependence on the coupling parameter \(\alpha\) as found for the electromagnetic field, one can infer the general features of the evaporation process. For \(\alpha=0\), the Hawking temperature is \(T_H(r_0)\propto r_0^{-1}\), the quantities \(\Xi_i(r_0)\) remain constant, and the black hole evaporates completely in a finite time. In contrast, for \(\alpha>0\), Eq.~(\ref{eq:eed}) admits a stable equilibrium point characterized by \(T_H(r_0)=0\), corresponding to the maximal allowed value of \(\alpha/r_0^2\). The horizon radius then approaches this remnant configuration only asymptotically, as the emission rate becomes exponentially suppressed~\cite{Dymnikova:2015yma}.

Our analysis indicates that the spacetime dimensionality has a direct impact on the rate of change of the event horizon. 
From Eq.~\eqref{eq:eed}, where we derived the relation for $\dot{r}_{0}$, we observe that for a fixed horizon radius $r_{0}$ the magnitude of $\dot{r}_{0}$ increases as the number of dimensions $D$ grows, suggesting that the dimensionality controls the evaporation rate. This behavior is consistently observed across all the regular black hole configurations considered in this work. In particular, we find that $|\dot{r}_{0}^{(5)}| < |\dot{r}_{0}^{(6)}|$ for all values of $\alpha$ within the allowed range. This result is in agreement with the qualitative expectations derived from the classical higher-dimensional singular solutions.

The suppression of evaporation in the regular configurations has a clear physical origin. It arises from the combined effect of the monotonic decrease of the Hawking temperature as $\alpha$ approaches its extremal value, and the modification of the effective potential, whose height and width increase with $\alpha$, thereby reducing the transmission coefficients. The former effect suppresses the thermal population of emitted quanta through the Bose--Einstein factor, while the latter reduces the probability of transmission across the potential barrier. The interplay of these two mechanisms leads to a systematic reduction of the total radiative power. Importantly, this behavior is robust across all models and dimensions analyzed here, indicating that the suppression is not tied to a specific metric choice but is instead a generic consequence of the regular core structure in quasi-topological gravity.

It is worth emphasizing that this mechanism qualitatively alters the late-time evaporation scenario. In contrast to the Schwarzschild-Tangherlini case, where the temperature increases as the mass decreases and evaporation accelerates toward a final explosive phase, the regular solutions exhibit a progressive cooling as the horizon shrinks. As $\alpha$ approaches its maximal value, the temperature tends to zero and the luminosity vanishes, causing the evolution to slow down asymptotically. The black hole therefore approaches a zero-temperature extremal configuration with finite mass determined by the coupling parameter $\alpha$, rather than evaporating completely. The dimensional dependence discussed above affects the rate at which this regime is approached, but does not change the qualitative endpoint of the evolution.

It is worth highlighting that the present analysis is restricted to the semiclassical regime. The regular black hole solutions considered here are treated as effective geometries valid within a semiclassical approximation, where quantum fields propagate on a fixed classical background. In this framework, the backreaction of the Hawking flux on the metric is incorporated only at the level of a quasi-static evolution of the horizon radius through Eq.~\eqref{eq:eed}, while a fully self-consistent solution of the semiclassical Einstein equations with $\langle T_{\mu\nu} \rangle$ is beyond the scope of the present work. The description remains valid as long as $r_0 \gg \ell_{\mathrm{Pl}}$, where quantum gravity effects can be safely neglected and the curvature scales remain well below the Planck scale. Near the Planck phase, or in the vicinity of the extremal remnant configuration where higher-curvature corrections may become dominant, a complete quantum gravity treatment would be required. Therefore, the results presented here should be understood as describing the semiclassical evaporation stage and the asymptotic approach toward the extremal configuration, rather than the ultimate microscopic fate of the remnant.

\section{Conclusions}\label{sec:conclusions}

In this work, we investigated the propagation of electromagnetic fields in a class of higher--curvature regular black hole spacetimes arising from quasi-topological gravity. These geometries, generated by infinite--curvature corrections, provide singularity-free black hole solutions while preserving a relatively simple analytic structure for static, spherically symmetric backgrounds. Our analysis focused on the impact of such corrections on the grey-body factors.

Our results show that, for all the regular black holes under consideration, the potential barrier becomes higher and broader, leading to a suppression of low-frequency transmission. Using the computed grey-body factors, we evaluated the energy emission rates for electromagnetic radiation. We found that increasing the coupling parameter $\alpha$ leads to a substantial suppression of the total radiated power. The emission spectra shift toward lower frequencies and decrease by several orders of magnitude as the coupling approaches its maximal allowed value. This behavior is well captured by an exponential fit dominated by the inverse Hawking temperature.

Although the present analysis is theoretical and confined to the semiclassical regime, the suppression of Hawking radiation and the emergence of extremal remnants may have potential phenomenological implications. In scenarios involving higher-dimensional gravity or TeV-scale fundamental Planck masses, such as certain braneworld or extra-dimensional models, the modified evaporation dynamics could affect the lifetime and observational signatures of microscopic black holes. Similarly, if primordial black holes formed in the early Universe are described by regular geometries, the suppression of late-time emission and the formation of long-lived remnants could alter their evaporation history and residual abundance, potentially impacting constraints derived from gamma-ray backgrounds or other cosmological probes. While a detailed phenomenological analysis lies beyond the scope of the present work, our results indicate that regular higher-curvature corrections may qualitatively modify black hole evaporation and therefore deserve consideration in studies connecting semiclassical gravity with observational signatures.

We also considered an effective evolution equation for the horizon radius during evaporation. Under the assumption that other perturbative channels exhibit a similar dependence on $\alpha$, our analysis suggests a qualitative modification of the evaporation process. While the Schwarzschild-like case $\alpha=0$ leads to complete evaporation in finite time, the presence of higher--curvature corrections results in an asymptotic approach to a remnant configuration characterized by a vanishing Hawking temperature and exponentially suppressed radiation.

An important outcome of the present analysis is that the suppression of Hawking radiation is universal across all regular black hole models (a)–(f) considered here and across all spacetime dimensions $D$. Although the quantitative magnitude of the emission rate depends on the particular field and the specific black hole model as well as on the dimensionality through the grey-body factors, multiplicity factors and the scaling of the Hawking temperature, the qualitative behavior is remarkably robust. As the coupling parameter $\alpha$ increases toward its maximal allowed value, both the Hawking temperature and the transmission coefficients decrease monotonically. As a result, the total emission rate is progressively suppressed and asymptotically vanishes in the extremal limit.

This behavior contrasts sharply with the Schwarzschild–Tangherlini solution of general relativity, for which the temperature diverges as the mass decreases, leading to complete evaporation in finite time. In all regular quasi-topological models studied here, by contrast, the evaporation process slows down continuously, and the black hole asymptotically approaches a zero-temperature extremal configuration with finite mass determined by the coupling parameter $\alpha$. As discussed in detail in Sec.~\ref{sec:energy}, this leads to a qualitatively different late-time evaporation scenario: instead of undergoing a final explosive phase, the system dynamically evolves toward a stable or long-lived remnant whose mass is fixed by the coupling parameter. The fact that this behavior persists across all models and dimensions considered indicates that the suppression of emission is not a feature of a particular metric choice, but rather a generic consequence of the regular core structure combined with the thermodynamic properties of quasi-topological gravity.

These results indicate that regular black holes may exhibit distinctive observational signatures, potentially allowing them to be distinguished from the singular black hole solutions of General Relativity. The suppression of grey-body factors and the associated modifications of the Hawking emission spectrum provide concrete imprints of the underlying higher--curvature structure of spacetime. Although the present analysis focused on electromagnetic fields as probes of these effects, graviton emission is expected to provide the dominant contribution in higher-dimensional spacetimes. A detailed study of gravitational perturbations and their corresponding emission channels therefore constitutes a natural extension of this work.

\acknowledgments
J. P. A. acknowledges support from the Conselho Nacional de Desenvolvimento Científico e Tecnológico (CNPq). The author is also grateful to Alexander Zhidenko for proposing the problem and reading the manuscript.

\bibliographystyle{JHEP}
\bibliography{bibliography.bib}

\end{document}